\documentclass[preprint,authoryear,12pt]{elsarticle}

\usepackage{epsfig}

\usepackage{amssymb}

\usepackage[ps2pdf,%
a4paper=true,%
breaklinks=true,%
colorlinks=true,%
pdfauthor={First Author et al.},%
pdftitle={Template for manuscripts in Advances in Space Research}%
]{hyperref}

\journal{Advances in Space Research}

\begin{document}

\begin{frontmatter}



\title{Stark broadening data for spectral lines
of rare-earth elements: Nb III\tnoteref{footnote1}}
\tnotetext[footnote1]{This paper is within the projects 176002, 176001 and III44022 of Ministry of Education, Science and Technological Development
of Republic of Serbia. The support of the French LABEX Plas@Par (UPMC, PRES Sorbonne Universties) is also acknowledged.
.}


\author{Zoran Simi\'c\corref{cor}}
\address{Astronomical Observatory, Volgina 7, 11060 Belgrade 38, Serbia}
\cortext[cor]{Corresponding author}
\ead{zsimic@aob.rs}


\author{Milan S. Dimitrijevi\'c}
\address{IHIS-Techoexperts, Be\v zanijska 23, 11080 Zemun, Serbia}
\address{Observatoire de Paris, 92195 Meudon Cedex, France}
\address{Astronomical Observatory, Volgina 7, 11060 Belgrade 38, Serbia}
\ead{mdimitrijevic@aob.rs}

\author{Luka \v C. Popovi\'c}
\address{Astronomical Observatory, Volgina 7, 11060 Belgrade 38, Serbia}
 \ead{lpopovic@aob.rs}

\begin{abstract}

The electron-impact widths for 15 doubly charged Nb ion lines
have been theoretically determined by using the modified semiempirical
method. Using the obtained results, we considered the influence of the electron-impact mechanism on line shapes
in spectra of chemically peculiar stars and white dwarfs.

\end{abstract}


\end{frontmatter}

\parindent=0.5 cm

\section{Introduction}

Spectral lines of rare earth elements (REE) are present
in stellar spectra, especially in spectra of
chemically peculiar (CP) ones and white dwarfs.
Consequently, data on the Stark broadening of REE spectral lines are
of interest not only for laboratory but also for astrophysical
plasma research as e.g. for REE abundance determination and opacity calculations.
The lines of Nb II are observed in spectra of cool Ap star $\gamma$ Equulei in the spectral region
$\lambda\lambda$ 3086-3807 \citep{Ade79}.
\citet{Fuh89} investigated eleven high resolution IUE spectra of the CP star H465
and reported presence of spectral lines for singly ionized
niobium among the other REE ions, which are quite strong in the optical spectra.

Atomic data for REE are needed in order to solve the astrophysical problems such as
the relative abundance of r- and s-process elements in metal-poor stars with enhanced neutron-capture
abundances. Slow neutron-capture process, the s-process occurs when the star is in the
"asymptotic giant branch" phase of its life. \citet{Nil10} derived
accurate transition probabilities for astrophysically interesting spectral lines of Nb II and Nb III and
determined the niobium abundance in the Sun and metal-poor stars rich in neutron-capture elements.
Also, \citet{Nil10} derived solar photospheric niobium abundance, obtaining the result log $\epsilon_{Sun} $ = 1.44 $\pm$ 0.06,
in agreement with the meteoritic value.

Due to very complex spectra of REE it is not always possible to apply the same method to calculate the Stark
broadening data (see, \citep{PD98}). Also, there is many cases with incomplete number of atomic data, when an
estimate of Stark broadening parameters on the basis of regularities and systematic trends can be possible \citep{DP89}.

In order to provide a complete set of Stark broadening data of astrophysical
interest, \citet{PD98} started and \citet{Pop99} continued a project to theoretically determine such data for a
number of spectral lines of REE.

\citet{Gay98} improved and extended the previous analyses of the Nb III spectrum on the basis of new,
high resolution, observations and modern technique of spectrum identification. The spectrum of doubly
charged Nb ion is recorded in the region 900-2512 \AA\- with the aid of a 6.65 m vacuum ultraviolet normal
incidence spectrograph. As a result of this work, the list of identified Nb III transitions contains
908 lines, including about 300 previously classified lines. \citet{Gay98} noted that LS-coupling is a quite
good approximation in the studied configurations for niobium. We note however, that there is many energy levels of  Nb III with
mixing of configurations.

Here, we have applied the modified semiempirical approach - MSE
\citep{DK80} whose applicability for complex spectra has been tested several times
(see e.g. \citep{PD96a,PD96b,PD97}), for the determination of electron-impac (Stark) full width at half maximum intensity (FWHM) of 15 Nb III spectral lines from 4d$^2$ ($^3$F)5s - 4d$^2$ ($^3$F)5p transitions. The obtained results will be used to investigate the importance of Stark broadening for plasma conditions in atmospheres of A type stars and DB white dwarfs.

\section{Theory}

For the determination of electron-impact line width Nb III lines, the modified semiempirical (MSE) approach \citep{DK80} has been used.
It is suitable, since in comparison with the semiclassical perturbation approach \citep{Sah69a,Sah69b},it needs a considerably smaller
number of atomic data, and, for Nb III, there is no a sufficient number of known atomic energy levels allowing more sophysticated
semiclassical perturbation calculations.
Within the MSE method, the electron-impact (Stark) full width (FHWM) of an isolated ion line, is given for an ionized emitter as:

$$
w_{MSE}=N{4\pi\over 3c}{\hbar^2\over m^2}\bigl({2m\over{\pi
kT}}\bigr)^{1/2}{\lambda^2\over\sqrt{3}}\cdot
\{\sum_{\ell_i \pm1}\sum_{L_{i'}J_{i'}}\vec{\Re}^2_{\ell_i,\ell_i\pm
1}\widetilde{g}(x_{\ell_i,\ell_i\pm
1})+
$$

$$
\sum_{\ell_f\pm1}\sum_{L_{f'}J_{f'}}\vec{\Re}^2_{\ell_f,\ell_f\pm
1}\widetilde{g}(x_{\ell_f,\ell_f\pm
1})
+\bigr(\sum_{{i'}}\vec{\Re}_{i{i'}}^2\bigl)_{\Delta n\ne
0}g(x_{n_i,n_i+1})+
$$

$$
\bigr(\sum_{{f'}}\vec{\Re}_{f{f'}}^2\bigl)_{\Delta n\ne
0}g(x_{n_f,n_f+1})\}, \eqno(1)$$

\noindent where the initial level is  denoted with $i$, the final one with
 $f$,  ${\vec{\Re}^2_{\ell_k,\ell_{k'}}, \ \ k=i,f}$ is the square of
the matrix element, and

$$\bigl(\sum_{k'}\vec{\Re}^2_{kk'}\bigr)_{\Delta n\neq 0}=
({3n^*_k\over{2Z}})^2{1\over 9}(n_k^{*2}+3\ell_k^2+3\ell_k+11), \eqno(2)$$
(in Coulomb approximation).

In Eq. (1)

$$x_{l_k,l_{k'}}={E\over{\Delta E_{l_k,l_{k'}}}}, k=i,f$$
where $E={3\over{2}}kT$ is the electron kinetic energy and $\Delta
E_{l_k, l_{k'}}=|E_{l{_k}}-E_{l_{k'}}|$ is the energy difference between levels
$l_k$ and $l_k$ $\pm$1 (k=i, f),

$$x_{n_k,n_k+1}\approx {E\over{\Delta E_{n_k,n_k+1}}},$$
where for $\Delta n\neq 0$ the energy difference between energy leves with
$n_k$ and $n_k$+1, $\Delta E_{n_k,n_k+1}$, is estimated as
 $\Delta E_{n_k,n_k+1}\approx 2Z^2E_H/n_k^{*3}$.
$n^*_k=[E_HZ^2/(E_{ion}-E_k)]^{1/2}$ is the effective principal quantum
number, $Z$ is the residual ionic charge, for example $Z$=1 for neutral atoms
and $E_{ion}$ is the appropriate spectral series limit.

In Eqs. (1) - (2)  $N$ and $T$ are  electron density and temperature,
respectively, while with $g(x)$ \citep{Gr68} and $\widetilde{g}(x)$ \citep{DK80} are denoted
 Gaunt factors for width, for $\Delta n \ne 0$ and  $\Delta n = 0$, respectively.

Atomic energy levels needed for calculation of Nb III Stark line widths have been taken from
\citet{Gay98}. In the Nb III spectrum configuration mixing is present so that exist terms which are a mixture of different configurations. If we can not neglect the contribution of other configurations, 
for any term represented as a mixture with K$_{1}$ part of the leading configuration and K$_{2}$ of the second one,  
where K$_{1}$ + K$_{2}$ = 1, we can use the expression:

$${\vec{\Re}^2_{j,j'}, }={\rm K_{1}}\vec{\Re}^2_{\alpha,\alpha'}+{\rm K_{2}}\vec{\Re}^2_{\beta,\beta'}$$

\noindent where ${\alpha,\alpha'}$ denote the energy level corresponding to the leading configuration, and its perturbing levels, and $\beta,\beta'$ is the same for the second configuration \citep{DP93}.

\section{Results and discussion}

Calculations of Stark widths for doubly charged niobium ion lines have been carried out for fifteen 4d$^2$ ($^3$F) 5s - 4d$^2$ ($^3$F)5p transitions.
We have selected all spectral lines from Nb III 5s-5p transition array where it is possible to apply MSE approach \citep{DK80}, and where the contributions of leading term of initial and final energy level are at least 80 per cent.

 In Table 1, our results for Stark full width for 4d$^2$ ($^3$F) 5s - 4d$^2$ ($^3$F)5p transitions of doubly charged niobium ion spectral lines, are shown.
The data are given for an electron density of $10^{17}$ cm$^{-3}$ and temperatures from
10,000 up to 300,000 K. Here, we draw attention to the 2415.2 \AA\ spectral lines of Nb III for transitions
5s $^{4}$F$_{7/2}$ - 5p $^{4}$F$^{o}_{7/2}$ with a relative intensity of 1000 \citep{Gay98}.

In order to show how much is important to take into account the Stark broadening for the modelling and
analysis of A type star and DB white dwarf atmospheres, we compared thermal Doppler and Stark broadening
in the both cases.

For A type stars, we used a model atmosphere with
$T_{eff}$  = 10,000 K and log $g$ = 4.5 \citep{Ku79}.
In Fig. 1, we compared thermal Doppler and Stark widths for Nb III spectral line
5s $^{4}$F$_{7/2}$ - 5p $^{4}$F$^{o}_{7/2}$ ($\lambda$=2415.2 \AA).
One can see, that towards the  subphotosheric
layers, Stark width becomes larger and in spite of the fact that it is smaller than Doppler one, in line wings its influence increases and may become of interest.

The influence of the Stark broadening on the same Nb III
spectral line (5s $^{4}$F$_{7/2}$ - 5p $^{4}$F$^{o}_{7/2}, \-\- \lambda$=2415.2 \AA)
has been tested also for DB white dwarfs, by using a model  with
$T_{eff}$ = 15,000 K and log $g$ = 8 \citep{Wi72}.
For the considered model atmosphere of the DB white dwarfs the prechosen
optical depth points at the standard wavelength $\lambda_{s}$=5150
\AA ($\tau_{5150}$) are used in \citet{Wi72} and here, as the
difference to the A type star model \citep{Ku79}, where the
Rosseland optical depth scale ($\tau_{Ross}$) has been taken.
In Fig. 2, the comparison of Stark and Doppler width as a function of optical depth, is shown for a model of DB white dwarf atmosphere.
Here, thermal Doppler broadening has much less importance in comparison
with the Stark broadening mechanism than in A-type stellar atmospheres.

The importance of the Stark broadening effect in stellar atmospheres under different plasma conditions
has been considered. Stark broadening may be not negligible in A type hot stars, and is the dominant in comparison with thermal Doppler in DB white dwarf atmospheres.
Also, Stark line widths for fifteen Nb III spectral lines, significant for synthesis and analysis of stellar spectra and for Nb abundance determination, have been determined.

\section{Figures}

\clearpage

\begin{figure}
\label{figure1}
\begin{center}
\includegraphics*[width=10cm,angle=0]{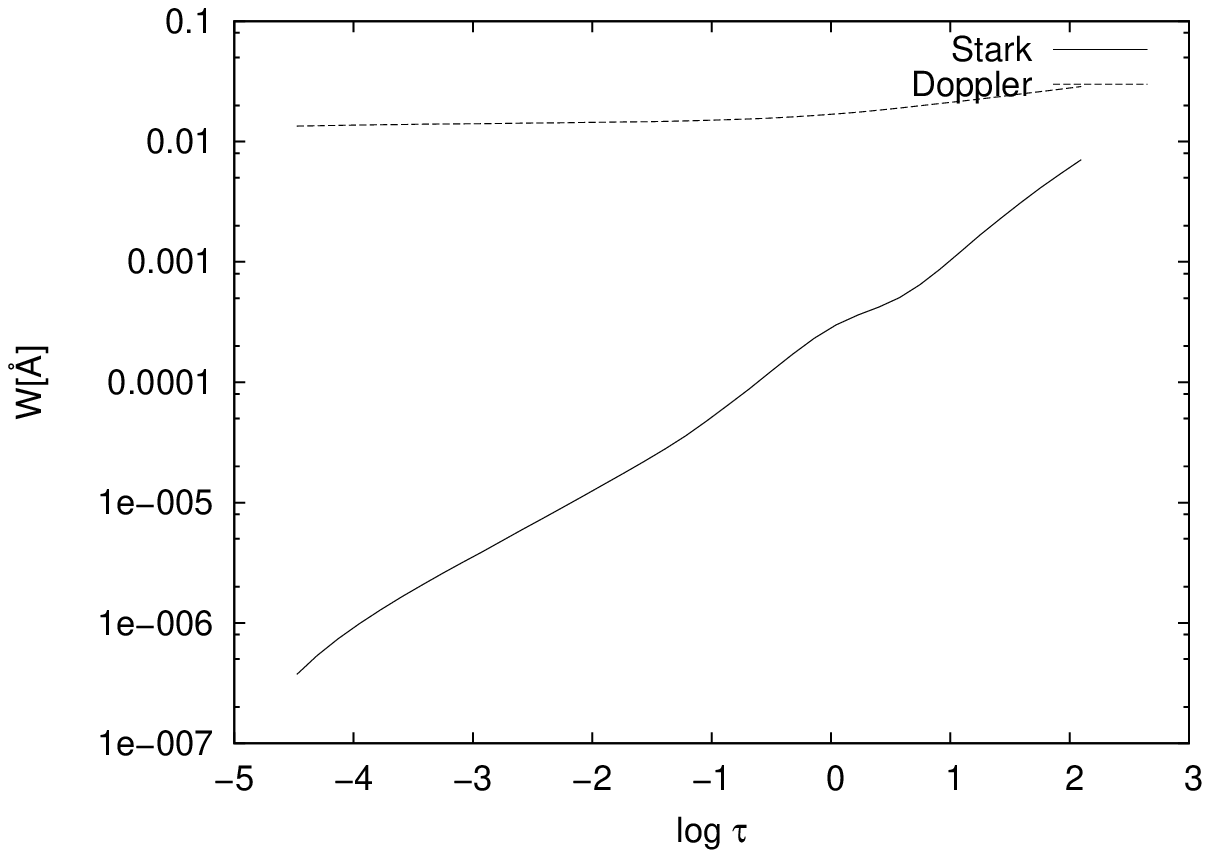}
\end{center}
\caption{Thermal Doppler and Stark widths for Nb III spectral lines
4d$^2$ ($^3$F) 5s $^{4}$F$_{7/2}$ - 4d$^2$ ($^3$F) 5p $^{4}$F$^{o}_{7/2}$ ($\lambda$=2415.2 \AA),
for an A type star atmosphere model with
$T_{eff}$  = 10,000 K and log $g$ = 4.5, as a function of the Rosseland
optical depth.}
\end{figure}

\begin{figure}
\label{figure2}
\begin{center}
\includegraphics*[width=10cm,angle=0]{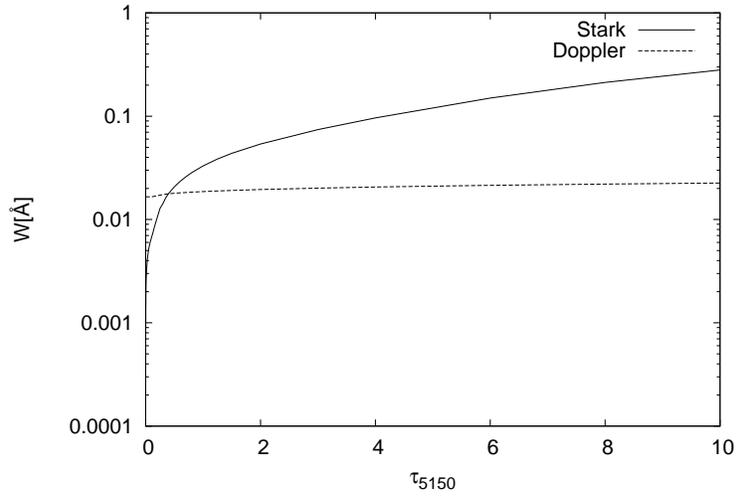}
\end{center}
\caption{Thermal Doppler and Stark widths for Nb III spectral lines
4d$^2$ ($^3$F) 5s $^{4}$F$_{7/2}$ - 4d$^2$ ($^3$F) 5p $^{4}$F$^{o}_{7/2}$ ($\lambda$=2415.2 \AA)
for a DB white dwarf atmosphere model with
$T_{eff}$  = 15,000 K and log $g$ = 8,
as a function of optical depth ${\tau}_{5150}$.} 
\end{figure}

\clearpage

\begin{table}
\caption{This table presents Nb III electron-impact broadening parameters (full
width at half maximum W) for 4d$^2$ ($^3$F) 5s - 4d$^2$ ($^3$F) 5p transitions obtained by the modified semiempirical method
\citep{DK80} for a perturber density of $10^{17}$ cm$^{-3}$ and temperatures from
10,000 up to 300,000 K.}

\begin{tabular}{llllll}
\hline
Transition & T(K) & W(\AA) & Transition & T(K) & W(\AA) \\
\hline
                                      &  10000.& 0.929-01&                                       &  10000.& 0.858-01\\
                                      &  20000.& 0.657-01&                                       &  20000.& 0.607-01\\
$^{4}$F$_{3/2}$ - $^{4}$G$^{o}_{5/2}$ &  50000.& 0.415-01& $^{4}$F$_{9/2}$ - $^{4}$F$^{o}_{7/2}$ &  50000.& 0.384-01\\
2599.7 \AA                            & 100000.& 0.302-01&           2469.5 \AA                  & 100000.& 0.278-01\\
                                      & 200000.& 0.249-01&                                       & 200000.& 0.228-01\\
                                      & 300000.& 0.238-01&                                       & 300000.& 0.219-01\\
\hline
                                      &  10000.& 0.736-01&                                       &  10000.& 0.823-01\\
                                      &  20000.& 0.521-01&                                       &  20000.& 0.582-01\\
$^{4}$F$_{3/2}$ - $^{4}$D$^{o}_{1/2}$ &  50000.& 0.329-01& $^{4}$F$_{9/2}$ - $^{4}$F$^{o}_{9/2}$ &  50000.& 0.368-01\\
2274.6 \AA                            & 100000.& 0.238-01&           2414.7 \AA                  & 100000.& 0.267-01\\
                                      & 200000.& 0.195-01&                                       & 200000.& 0.218-01\\
                                      & 300000.& 0.188-01&                                       & 300000.& 0.210-01\\
\hline
                                      &  10000.& 0.988-01&                                       &  10000.& 0.956-01\\
                                      &  20000.& 0.698-01&                                       &  20000.& 0.676-01\\
$^{4}$F$_{9/2}$ - $^{4}$G$^{o}_{7/2}$ &  50000.& 0.442-01& $^{4}$F$_{5/2}$ - $^{4}$G$^{o}_{5/2}$ &  50000.& 0.427-01\\
2657.3 \AA                            & 100000.& 0.321-01&           2635.0 \AA                  & 100000.& 0.311-01\\
                                      & 200000.& 0.264-01&                                       & 200000.& 0.256-01\\
                                      & 300000.& 0.253-01&                                       & 300000.& 0.245-01\\
\hline
                                      &  10000.& 0.924-01&                                       &  10000.& 0.901-01\\
                                      &  20000.& 0.653-01&                                       &  20000.& 0.637-01\\
$^{4}$F$_{9/2}$ - $^{4}$G$^{o}_{9/2}$ &  50000.& 0.413-01& $^{4}$F$_{5/2}$ - $^{4}$G$^{o}_{7/2}$ &  50000.& 0.403-01\\
2558.7 \AA                            & 100000.& 0.300-01&           2546.4 \AA                  & 100000.& 0.293-01\\
                                      & 200000.& 0.246-01&                                       & 200000.& 0.241-01\\
                                      & 300000.& 0.237-01&                                       & 300000.& 0.231-01\\
\hline
                                      &  10000.& 0.859-01&                                       &  10000.& 0.788-01\\
                                      &  20000.& 0.608-01&                                       &  20000.& 0.557-01\\
$^{4}$F$_{9/2}$ - $^{4}$G$^{o}_{11/2}$ &  50000.& 0.384-01& $^{4}$F$_{5/2}$ - $^{4}$F$^{o}_{7/2}$ &  50000.& 0.352-01\\
2457.8 \AA                            & 100000.& 0.279-01&           2373.5 \AA                  & 100000.& 0.256-01\\
                                      & 200000.& 0.229-01&                                       & 200000.& 0.209-01\\
                                      & 300000.& 0.220-01&                                       & 300000.& 0.202-01\\
\hline
\end{tabular}
\label{table1}
\end{table}

\clearpage

\addtocounter{table}{-1}
\begin{table}
\caption{Continued}

\begin{tabular}{llllll}
\hline
Transition & T(K) & W(\AA) & Transition & T(K) & W(\AA) \\
\hline
                                      &  10000.& 0.996-01&                                       &  10000.& 0.818-01\\
                                      &  20000.& 0.704-01&                                       &  20000.& 0.579-01\\
$^{4}$F$_{7/2}$ - $^{4}$G$^{o}_{5/2}$ &  50000.& 0.445-01& $^{4}$F$_{7/2}$ - $^{4}$F$^{o}_{7/2}$ &  50000.& 0.366-01\\
          2686.5 \AA                  & 100000.& 0.324-01&           2415.2 \AA                  & 100000.& 0.265-01\\
                                      & 200000.& 0.266-01&                                       & 200000.& 0.217-01\\
                                      & 300000.& 0.255-01&                                       & 300000.& 0.209-01\\
\hline
                                      &  10000.& 0.938-01&                                       &  10000.& 0.785-01\\
                                      &  20000.& 0.664-01&                                       &  20000.& 0.555-01\\
$^{4}$F$_{7/2}$ - $^{4}$G$^{o}_{7/2}$ &  50000.& 0.420-01& $^{4}$F$_{7/2}$ - $^{4}$F$^{o}_{9/2}$ &  50000.& 0.351-01\\
          2594.5 \AA                  & 100000.& 0.305-01&           2362.8 \AA                  & 100000.& 0.255-01\\
                                      & 200000.& 0.251-01&                                       & 200000.& 0.209-01\\
                                      & 300000.& 0.241-01&                                       & 300000.& 0.201-01\\
\hline
                                      &  10000.& 0.879-01& & & \\
                                      &  20000.& 0.662-01& & & \\
$^{4}$F$_{7/2}$ - $^{4}$G$^{o}_{9/2}$ &  50000.& 0.393-01& & & \\
          2500.5 \AA                  & 100000.& 0.286-01& & & \\
                                      & 200000.& 0.235-01& & & \\
                                      & 300000.& 0.226-01& & & \\
\hline
\end{tabular}
\label{table1}
\end{table}

\end{document}